 \documentclass[12pt,preprint]{aastex}
 
\newcommand{\etal}{\mbox{et~al.}}

\def\deg      {{\ifmmode^\circ\else$^\circ$\fi}} 

 \shorttitle{Faint End Slopes Of Galaxy Luminosity Functions}
 \shortauthors{Liu et al.}
 
\begin{document}
  
 \title{The Faint End Slopes Of Galaxy Luminosity Functions 
In The COSMOS 2-Square Degree Field\altaffilmark{*}}
 
\author{ 
Charles T. Liu\altaffilmark{1},
Peter Capak\altaffilmark{2},
Bahram Mobasher\altaffilmark{3},
Timothy A. D. Paglione\altaffilmark{4},
R. Michael Rich\altaffilmark{5},
Nicholas Z. Scoville\altaffilmark{2},
Shana M. Tribiano\altaffilmark{6}, and
Neil D. Tyson\altaffilmark{7}
}
 
\altaffiltext{$\star$}{Based on observations with the NASA/ESA {\em
Hubble Space Telescope}, obtained at the Space Telescope Science
Institute, which is operated by AURA Inc, under NASA contract NAS
5-26555; also based on data collected at: 
Kitt Peak National Observatory, Cerro Tololo Inter-American Observatory, 
and the National Optical Astronomy Observatory, which are
operated by the Association of Universities for Research in Astronomy, Inc.
(AURA) under cooperative agreement with the National Science Foundation; 
the Subaru Telescope, which is operated by the National Astronomical Observatory of Japan;
the XMM-Newton, an ESA science mission with instruments and 
contributions directly funded by ESA Member States and NASA; 
the European Southern Observatory under Large Program 175.A-0839, Chile; 
the Canada-France-Hawaii Telescope with MegaPrime/MegaCam operated as a
joint project by the CFHT Corporation, CEA/DAPNIA, the National Research
Council of Canada, the Canadian Astronomy Data Centre, the Centre National
de la Recherche Scientifique de France, TERAPIX and the University of Hawaii; and
the National Radio Astronomy Observatory, which is a facility of the National Science 
Foundation operated under cooperative agreement by AURA.}  
\altaffiltext{1}{Astrophysical Observatory, Dept. of Engineering Science and Physics, City University of New York, College of Staten Island, 2800 Victory Blvd, Staten Island, NY  10314}
\altaffiltext{2}{California Institute of Technology, MC 105-24, 1200 East
California Boulevard, Pasadena, CA 91125}
\altaffiltext{3}{Space Telescope Science Institute, 3700 San Martin
Drive, Baltimore, MD 21218}
\altaffiltext{4}{City University of New York, York College, 94-20 Guy R. Brewer Blvd., Jamaica, NY  11451}
\altaffiltext{5}{Department of Physics and Astronomy, University of
California, Los Angeles, CA 90095}
\altaffiltext{6}{City  University of New York, Borough of Manhattan Community College, 199 Chambers St., New York, NY 10007}
\altaffiltext{7}{American Museum of Natural History, Central Park West at 79th Street, New York, NY  10024}
  
 \begin{abstract}
 
We examine the faint-end slope of the rest-frame V-band luminosity function (LF), 
with respect to galaxy spectral type, of field galaxies with 
redshift $z < 0.5$, using a sample of 80,820 galaxies with
photometric redshifts in the Cosmic Evolution Survey (COSMOS) field.  
For all galaxy spectral types combined, the LF slope ranges 
from $-1.24$ to $-1.12$, from the lowest redshift bin to the highest. 
In the lowest redshift bin ($0.02 < z < 0.1$), where the magnitude limit is
$M_{V} \lesssim -13$, the slope ranges from $\alpha \sim -1.1$ for
galaxies with early-type spectral energy distributions (SEDs), to
$\alpha \sim -1.9$ for galaxies with low-extinction starburst SEDs.
In each galaxy SED category (Ell, Sbc, Scd/Irr, and starburst), the
faint-end slopes grow shallower with increasing redshift; in the highest
redshift bin ($0.4 < z < 0.5$),
$\alpha \sim -0.5$ and $-1.3$ for early-types and starbursts respectively.
 The steepness of $\alpha$ at
lower redshift could be qualitatively explained by large numbers of faint
dwarf galaxies, perhaps of low surface brightness, which are not detected
at higher redshifts.

\end{abstract}

 
 \section{Introduction}
 
The luminosity function (LF) of galaxies varies substantially with 
respect to many key physical parameters such as galaxy morphology, 
environment, color, star formation rate,
surface brightness, and redshift.  These many differences serve as 
powerful diagnostics of the broad tapestry of galaxy evolution.  
 
The most important requirement for the accurate
derivation of galaxy LFs are large, complete samples of galaxies
with reliable photometry and redshift determinations.  The Cosmic 
Evolution Survey (COSMOS; \citealt{sco07a}) contains not only the 
largest contiguous area of the sky yet observed with the 
Hubble Space Telescope, but also deep multiwavelength
imaging across the entire 2 \sq\deg COSMOS field
(\citealt{cap07}).  COSMOS thus affords deep, homogeneous 
photometric and photometric redshift coverage for a sample
of  some $10^6$ galaxies -- complementing well at higher redshifts
the largest galaxy surveys of the relatively nearby 
universe with which galaxy LFs have been derived, such as the Two-Degree
Field Galaxy Redshift Survey (2dFGRS; \citealt{col01,cro05}) 
and the Sloan Digital Sky Survey (SDSS; \citealt{bla03,aba04}).

Comprehensive analyses of the LF characteristics 
of the entire COSMOS galaxy sample
will ultimately be forthcoming, with the completion 
of the spectroscopic portion of the survey
(\citealt{lil07}) and the continued addition and refinement of the 
photometric and photometric redshift measurements in multiple bandpasses
(\citealt{cap07,mob07}).  Already, though, it is feasible to address 
important scientific questions about galaxy LFs with the current
optical and near-infrared multiband data (\citealt{cap07,tan07,aji07,sca07})
in conjunction with the Hubble Space Telescope $I_{814}$ broad-band 
ACS imaging (\citealt{sco07b}).
 
One particular component of the field galaxy LF, the faint end ($M_{AB} > -18$)
of the LF at low to intermediate ($0 \lesssim z \lesssim 0.5$) redshifts, 
is uniquely well-suited for analysis with the COSMOS data.  This population
of galaxies lies in an apparent magnitude range somewhat beyond the 2dFGRS
and SDSS survey limits; they are too faint for spectroscopic observations, with
redshifts too low to be included in surveys optimized for high-redshift galaxies.
Yet, exploring these galaxies' contribution to the overall LF is critical
for understanding galaxy evolution during the latter half of cosmic history.

Recent comprehensive studies of field galaxy LF evolution have focused on the luminosity 
evolution of the bright end ($M_{AB} \lesssim -20$).  There, some consensus 
appears to be gradually emerging about the extent of that evolution.  
\citet{dah05}, \citet{wil06}, and \citet{sca07} all find a brightening of $M_{B}*$ by
$\sim 1$ magnitude between $0 \lesssim z \lesssim 1$.  In shorter-wavelength 
bandpasses, the luminosity evolution is more pronounced, while at longer
wavelengths it appears to be weaker and possibly even in the negative sense,
dimming in the near-infrared $J$-band from $z \sim 0.4$ to $0.9$ (\citealt{dah05}).
  
The evolution of the faint-end slope, however, remains highly uncertain.  
At low redshifts, results from the Sloan Digital Sky Survey 
(\citealt{bla05,bal05}), the 2dF Galaxy Redshift Survey (\citealt{cro05}), and
other datasets such as that from the Century Survey (\citealt{bro01}) and the GALEX mission (\citealt{bud05}) roughly agree, for example, on a moderate slope of $\alpha \sim -1.1$.
Beyond redshifts of a few tenths, the faint-end slope becomes very difficult to
address, mainly because the number of low-luminosity galaxies detected 
in galaxy surveys decreases dramatically with increasing redshift.  
Despite a number of efforts to measure
evolution in the faint-end slope at redshifts less than $z \sim 1$ 
(\citealt{wol03,ilb05,zuc06}), very little is known for galaxies fainter 
than $M \sim -18$.  This is in part because the relationship between
the bright-end and faint-end characteristics of galaxy LFs is not straightforward, 
often resulting in a tradeoff between the precision
LF evolution measurements at the two ends.  \citet{bal05} and 
\citet{wil06}, for example, each select fixed LF faint-end slopes based on
galaxies brighter than $M \sim -18$, and use that constraint
throughout their bright-end evolution measurements.

Whether or not the faint-end slope evolves with redshift, however, 
it is clear that its steepness varies widely for galaxies of different morphological 
and spectral types, indicating substantial differences in the evolutionary
histories of galaxies.  Broadly speaking, irregular galaxies,
"blue" galaxies, and strongly star-forming galaxies -- three significantly
overlapping galaxy sub-populations -- evolve more strongly than galaxies
of other types.  Such galaxies are characterized by very steep faint-end LF 
slopes and substantial
evolution in luminosity and/or number density at even moderate redshifts
of $z \lesssim 0.5$ (\citealt{mar94,lil95,ell96,liu98b,bro98,lin99}).  More recent work
has further confirmed and quantified this trend at higher redshifts (see, 
e.g., \citealt{che03,per05,dah05}).    At low redshifts, evidence is mounting that
composite parameterizations may more accurately reflect the shape of the LF than 
the usual single-function ones
(\citealt{del03,del04,bla05}); and that faintward of $M \sim -18$, the 
power-law slope of the LF may differ quantitatively from the slope 
brightward of that threshold (\citealt{mad02,nor02,bla05}).  
This could arise, for example, due to large numbers of low-luminosity "blue" 
galaxies (\citealt{wol03}) or low surface brightness galaxies (\citealt{imp96,bla05})
which may have previously evaded detection.  

In this paper, we present measurements of the 
faint-end slopes of the rest-frame V-band luminosity functions of
galaxies in the COSMOS survey at
$0 \lesssim z \lesssim 0.5$, focusing in particular on the 
change in that slope as a function 
of redshift and galaxy spectral type.  
The depth and breadth of the COSMOS multiband photometry allows for 
reliable identifications of galaxy redshift and spectral type, with robust
redshift error estimates for each galaxy, to a limit of $m_{AB} \sim 25$
in the optical passbands.  Even so, the relatively large 
and varying redshift uncertainties of photometric redshifts
can present substantial quantitative challenges and systematic biases
(\citealt{sub96,liu98b,che03,dah05}).  We use Monte Carlo simulations 
to characterize these biases, and to recover the faint-end slopes of
galaxy-type specific LFs.  

This work represents an initial study of the general properties of these
type-specific faint-end LF slopes, to provide quantitative comparisons with the 
results of other large field galaxy surveys.  Studies of the bright-end evolution of 
the galaxy LF from the COSMOS survey are given elsewhere 
(e.g., \citealt{sca07}); and a more detailed breakdown of the 
$0 \lesssim z \lesssim 0.5$ galaxy population by redshift,
galaxy spectral type, and galaxy morphology will be presented
in a future paper (Liu et al., in preparation).

Throughout this paper, we adopt a flat cosmology
with $\Omega_{\Lambda} = 0.7$, $\Omega_{m} = 0.3$, and
$H_{o} = 70$ km s$^{-1}$  Mpc$^{-1}$.

\section{Galaxy Sample and Photometric Redshifts}

In our analysis, we use a compilation of the COSMOS optical/near-infrared
data (\citealt{cap07,mob07}), which includes observations
with HST Advanced Camera for Surveys ($I_{814}$), 
the Subaru Telescope ($B, V, r', i', z', NB816$), the Canada-France-Hawaii
Telescope ($u*, i*$), and the 4-meter KPNO Mayall and
CTIO Blanco Telescopes ($K_{s}$), as well as supplementary
data from the Sloan Digital Sky Survey.  The data from the different
telescopes were all matched to a common pixel scale and
smoothed to the same point source function.
SExtractor (\citealt{ber96}) was then used in dual mode to generate a photometric
catalog, selected using the Subaru $i'$ and CFHT $i*$ images.   
The limiting 3$\sigma$ AB magnitude is $i' = 26.03$.
A detailed description of the imaging data, photometry, and
photometric calibration are given in \citet{cap07}.

\begin{figure}
\epsscale{1.0}
\plotone{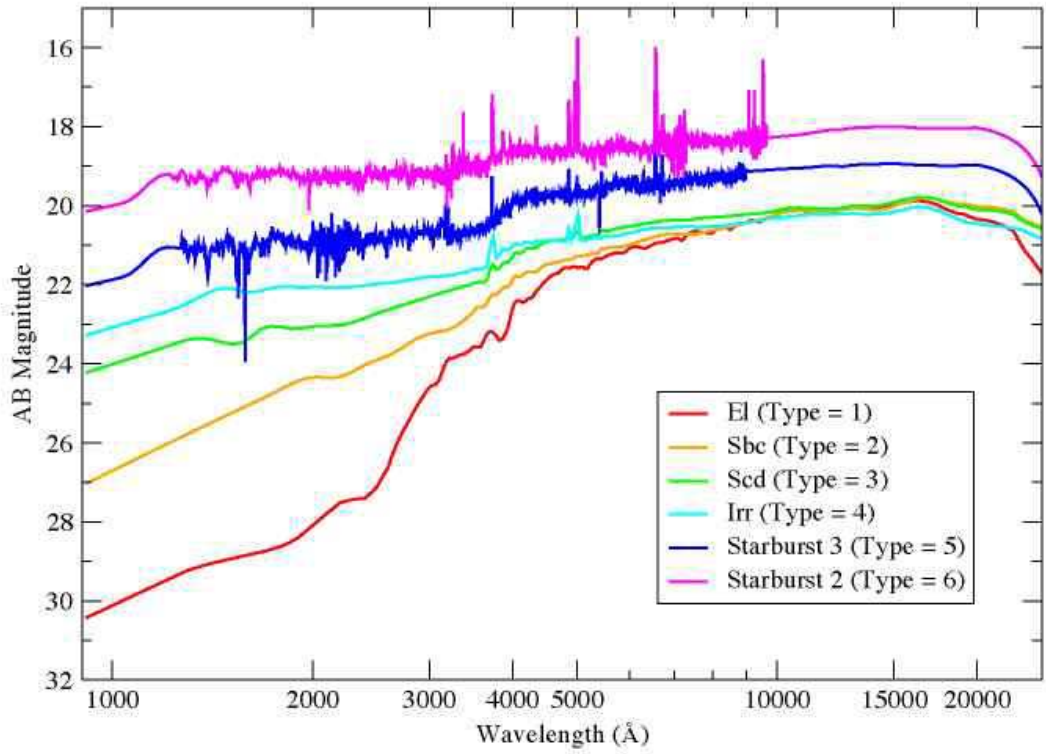}
\caption{
Spectral energy distributions from \citet{mob07} used to compute COSMOS 
galaxy spectral types and photometric redshifts.
}
\label{SEDs}
\end{figure}

Photometric redshifts for individual galaxies were computed using 
the methods described in \citet{mob07}.  
Six basic galaxy spectral types, adapted from the four template 
types -- E, Sbc, Scd, and Im -- from
\citet{col80}, and the starburst templates SB2 and SB3 of
\citet{kin96}, were used.  These templates are presented graphically 
in Figure \ref{SEDs}.  These galaxy spectral types are derived 
from empirical data, and represent the range of non-AGN galaxy 
spectral energy distributions (SEDs) from redder to bluer colors; 
these starburst SEDs, for example, represent very blue galaxies 
not significantly reddened or obscured by dust.  Interpolation was used between the six 
spectral types to produce a grid of 31 possible galaxy SED fits.    

Using the multiband photometry, the COSMOS photometric redshift code 
(\citealt{mob07}) was used to derive a
photometric redshift $z_p$ for each galaxy, with 68\% and 95\% confidence 
intervals computed above and below that value.  To determine the accuracy of
the code, $z_{p}$ values were compared with spectroscopic
redshifts ($z_{s}$) in 868 galaxies with $z < 1.2$ and $i_{AB} \lesssim 24$ 
with secure redshift measurements from the {\it{z}}COSMOS survey (\citealt{lil07}).
The galaxy spectral types (20\% early-type, 63\% spiral, 17\% starburst) 
are evenly distributed with redshift in the spectroscopic sample, and about 
half of the sample (45\%) is at $z_s \leq 0.5$.

The detailed statistics of the $z_p$ and $z_s$ comparisons are described in
\citet{mob07}, primarily in terms of the 
parameter $\Delta z = (z_p - z_s) / (1 + z_s)$.  As given in
Table 4 and Figure 5 of \citet{mob07},
the nominal dispersion between these photometric
redshifts and spectroscopically determined redshifts was
rms ($\Delta z$) $= 0.033$ for non-AGN galaxies.  
With respect to the different spectral types, 
\citet{mob07} showed that early-type, spiral, and starburst galaxies
have rms $(\Delta z) =$ 0.034, 0.030, and 0.042 respectively.

\begin{figure}
\epsscale{1.0}
\plotone{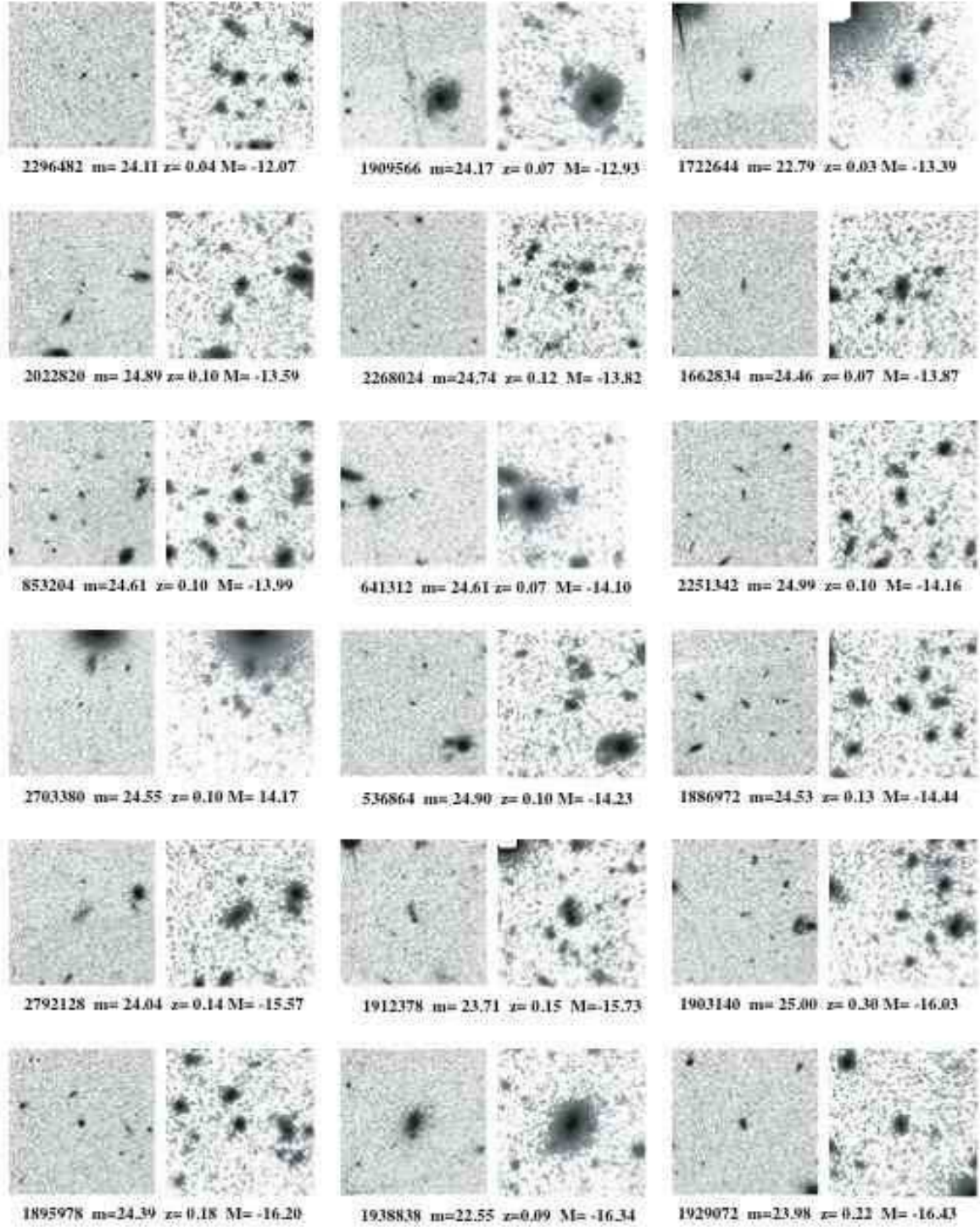}
\caption{
Hubble $ACS$  $I_{814}$ and Subaru $B$ images of a representative
selection of very low-luminosity galaxies in the COSMOS field.  Each image is
$15"$ across, and each image pair is labeled by the COSMOS catalog ID 
number, photometric redshift, and apparent and absolute $V$ magnitudes
of the object at the image center.
}
\label{IMGs}
\end{figure}

We chose to compute our luminosity functions using the $V$-band 
data for this study.  The saturation limit for bright objects in that
image was $V \sim 18.5$, and the 3$\sigma$ faint-detection limit
was $V \sim 26.4$; so we conservatively chose as our 
apparent magnitude range $19.0 < m_{V} < 25.0$.  
Details of the photometry are described in \citet{cap07}.
A rest-frame absolute $M_{V}$ magnitude,
based on the derived $z_p$ and using the
K-correction for the best-fit spectral type, was
computed for each galaxy.  Ground and space-based
images of a representative subsample of the faintest galaxies 
($M_{V} \sim -16$ and fainter)
in this sample are presented in Figure \ref{IMGs}.

\section{Luminosity Functions}

As with all surveys which rely primarily on photometric rather than
spectroscopic redshifts, the application of the COSMOS galaxy 
sample to the derivation of galaxy LFs requires great care to account
for both random and systematic errors in the redshift determinations.
Just as methods of computing photometric redshifts have evolved and
improved (\citealt{koo86,con95,liu98a,ben00,mob04,mob07}), so too have 
techniques to quantify and compensate for the effects of relatively large 
redshift error bars in LF calculations (\citealt{sub96,liu98b,che03,dah05,per05}).

In this work, we adapt the method used in \citet{liu98b}, updating it with
additional components similar to those used in more recent studies
(e.g., \citealt{che03,dah05,per05})
to reproduce the faint-end slope of the LFs of COSMOS galaxies.
Our strategy is based on the 1/$V_{max}$ method (\citealt{sch86}), which is
well described by \citet{che03} as a maximum likelihood method for estimating a luminosity function without assuming any parametric form.  We account for photometric redshift errors by weighting the galaxies as probability-smoothed luminosity distributions at the redshifts where they are measured.

\subsection{ The modified 1/$V_{max}$ method }

Consider a galaxy with an apparent magnitude $m_f$ in a passband
$f$, and redshift
$z \pm \sigma$.  If $\sigma$=0, then the absolute magnitude is 

\centerline{ $M_f$ = $m_f -$ 5 log(d$_L(z)$) $-$ 25.0 $- k_{f}(z)$ }

\noindent
where d$_L(z)$ is the luminosity distance in Megaparsecs, and
$k(z)$ is the K-correction at that redshift, in that passband, for the spectral 
energy distribution of the galaxy.  The contribution of that
galaxy to the luminosity distribution is then a delta 
function of amplitude unity at redshift $z$.  

In the case where $\sigma >$ 0,
and the error distribution is Gaussian, the galaxy
can be thought of as adding a series of
fractional contributions to the luminosity distribution
in the redshift space surrounding $z$.  Such a fraction at,
for example, redshift $z + \delta z$ and with a
differential redshift width $dz$, would have an absolute magnitude

\centerline{ $M_f^{\arcmin}$ = $m_f -$ 5 log(d$_L(z + \delta z)$) $-$ 25.0 $- k_{f}(z + \delta z)$ }

\noindent
and have an amplitude

\centerline{ N$_{z + \delta z}$ = 
P$_G(z + \delta z,z,\sigma) dz$ / A$_G(z + \delta z,z,\sigma)$}

\noindent
where P$_G$ and A$_G$ are the Gaussian probability function and its integral,
respectively (see, e.g., \citealt{bev92}).  

For the photometric redshifts of the COSMOS survey, the redshift error distribution is not Gaussian, but rather can be modeled as two half-Gaussians (\citealt{cap07,mob07}), where the
68\% confidence interval on the lower 
and upper limits are $\sigma_{l}$ and $\sigma_{u}$ respectively.  For a galaxy 
which a photometric redshift $z_p$, the amplitude of the luminosity contribution function would be

\centerline{ N$_{z + \delta z}$ = 
P$_G(z + \delta z,z,\sigma_{l}) dz$ / A$_G(z + \delta z,z,\sigma_{l})$ } 

\noindent
for  $z < z_p$, and

\centerline{ N$_{z + \delta z}$ = 
P$_G(z + \delta z,z,\sigma_{u}) dz$ / A$_G(z + \delta z,z,\sigma_{u})$} 

\noindent
for $z  > z_p$.

This ``fuzzing'' of a galaxy's luminosity distribution 
in redshift space is straightforwardly achieved 
numerically, with a choice of $dz << \sigma$ to minimize random magnitude
errors.  For this COSMOS dataset, we used $dz_l$ = 0.02 $\sigma_l$ and 
$dz_u$ = 0.02 $\sigma_u$.  This divides each galaxy into a Gaussian-weighted
luminosity distribution with 300 bins, from $z - $3$\sigma_l$ to
$z + $3$\sigma_u$.  The entire distribution for each 
galaxy is normalized to unity.  

In the standard 1/V$_{max}$ method, each galaxy contributes a weight
to the luminosity function equal to the inverse of the accessible
volume within which it can be observed.  The accessible volume,
referred to here as V$_{max}$, is the total comoving volume within
the redshift boundaries of the sample, where the given galaxy could
be and fall within the selection criteria of the sample.  In our case,
the relevant criteria are the bright and faint apparent magnitude limits,
and the effective solid angle of the COSMOS survey.

In the case of a probability-weighted luminosity distribution for
individual objects, it is straightforward to compute V$_{max}$ for 
each fractional galaxy;
correspondingly, its contribution to the luminosity
function is (1/V$_{max}$) $\times$ N$_{z + \delta z}$.  Assembling
the luminosity function is then a matter of summing those contributions
within absolute magnitude bins.

\subsection{Redshift Limits And Sample Size}

Since the primary goal of this work was to examine the LF faint-end slope,
the upper redshift boundary was determined mainly by our
desire to sample with high completeness at least as faint as 
$M_{V} \sim -16.5$ in the entire redshift range.  For a typical galaxy in
the sample, depending on the galaxy's K-correction, this corresponds 
roughly to $z \lesssim 0.4$.  But because each galaxy's luminosity 
is calculated as a probability-weighted distribution, there is a statistically
significant contribution to the LF more than a full magnitude beyond the
formal absolute magnitude limit.  Thus, with the caveat that we are beyond
that limit, we were also able to derive faint-end slopes of 
the LFs of galaxies in the redshift range $0.4 < z \leq 0.5$.

Similarly, at the low redshift range, we are also able to measure fainter 
in absolute magnitude than the formally faintest detectable galaxy.   However,
to avoid large systematic magnitude errors and biases from structure in the local universe, we set a lower redshift bound of $z > 0.02$ for deriving the LFs.
This meant that we have statistically meaningful luminosity contributions to
the LFs down to an absolute magnitude of $M_{V} \sim -12.8$ for $0.02 < z \leq 0.1$.   

Within our apparent magnitude limits of $19.0 < V < 25.0$, the 
COSMOS survey photometric and $z_p$ catalog (\citealt{cap07})
contains 49,161 galaxies in the redshift range $0.02 < z \leq 0.5$.  Below
and above this redshift range, however, there are galaxies within the
apparent magnitude limits whose 
probability-smoothed luminosity distributions contribute
to the light within that range.  Using $\delta_{68}$ to denote the width of
the 68\% confidence interval for $z_p$, we thus also include the contributions
of all other galaxies whose luminosity distribution tails fall within $3 \times \delta_{68}$.
(For example, a galaxy with $z_p = 0.51$ and $\delta_{68} = 0.05$
would contribute the portion of its luminosity distribution from 
$0.36 \leq z \leq 0.50$, while a galaxy with $z_p = 0.61$ and 
$\delta_{68} = 0.05$ would contribute from $0.46 \leq z \leq 0.50$.)
There are 31,659 galaxies in the catalog that make such a partial contribution;
even though many of those galaxies add only a tiny fraction of a galaxy into
the $z \leq 0.5$ redshift range, we
include them in our analysis for statistical completeness.  Thus, a total
of 80,820 galaxies are included in the sample used to derive the LFs.

To check how the inclusion of these galaxies might affect the distribution of
$z_p$ accuracy in the sample as a whole, we created
a subset of the 80,820-galaxy sample where the galaxies have a
$\delta_{68}$ of at most 0.1.
This subset of 41,237 galaxies (51\% of the entire sample)
effectively contains objects whose $\delta_{68}$ are
no more than 3 times the rms $(\Delta z)$ of the spectroscopically
tested accuracy of the COSMOS $z_p$ code.  We then computed the 
mean, median, and rms of the $\delta_{68}/(1+z)$ values for the
41,237, 49,161, and 80,820 galaxy samples.  We give the results
in Table 1.  As might be expected, the median, mean, and rms values all increase
as the constraints in redshift and $\delta_{68}$ are lifted and the sample sizes grow.
The median $\delta_{68}/(1+z)$ value increases only modestly on an absolute
numerical basis (0.042 to 0.043 to 0.052); the mean and rms increase more
substantially.

\subsection{Simulations}

Computing LFs using "fuzzy" galaxies with photometric redshifts
is clearly vulnerable to a set of systematic
errors that would not be present for a galaxy sample with secure spectroscopic
redshifts.  In this work, we use the standard galaxy LF parameterization of 
\citet{sch76}, where $M^{*}$ is the characteristic 
magnitude, $\phi^{*}$ the characteristic number density, and $\alpha$ the 
faint-end slope (\citealt{lin96}):

\begin{equation}
\phi(M) =  \phi^{\ast}  \times 0.4 \ln 10 
  \exp\left(-10^{-0.4(M-M_{\ast})}\right) 
  \left[10^{-0.4 \left( M-M_{\ast} \right)(1 + \alpha)} 
  \right] dM
\end{equation}

In any given redshift bin, objects near the peak of the LF --- that is, near
$M*$ --- will have part of their light distributed toward brighter magnitudes;
and objects at the bright and faint end of the galaxy sample will have
their light scattered still further.  This would cause an overestimate of those parts
of the LF which contribute the least light.  Additionally, as the
luminosity contributions of the galaxies are distributed across a large
redshift range, light from galaxies inside a given redshift bin will sometimes
be scattered out of that bin; and since lower-$z$ redshift bins have smaller
volumes than higher-$z$ ones, there is the risk that more light would be added into
lower-redshift bins than would be removed from them.  This could bias the
distribution of derived absolute magnitudes in those bins.

We quantify and correct for these errors using Monte Carlo simulations in the
manner described by \citet{liu98b}.  
For an arbitrary fixed $M*$ and $\phi^{*}$,
we created populations of galaxies that followed Schechter functions
with faint-end slopes ranging from $-0.2 < \alpha < -2.2$.  
For each $\alpha$, we populated a simulated COSMOS survey volume with 
the corresponding galaxy population.  We then "detected" these galaxies based on
the magnitude and redshift limits of our survey; and for each galaxy detected,
we randomly added an error to the redshift of that galaxy consistent with the 
measured dispersion of the COSMOS photometric redshift catalog, 
rms ($\Delta z) = 0.033$.  Finally, we assigned to that galaxy a $\delta_{68}$ equal to the median 
$\delta_{68}$ of the 80,820-galaxy sample, i.e. $0.052 \times (1+z_{s})$. 
The LF for this simulated galaxy sample was then computed using the 
modified $1/V_{max}$ method described above.  

\begin{figure}
\epsscale{1.0}
\plotone{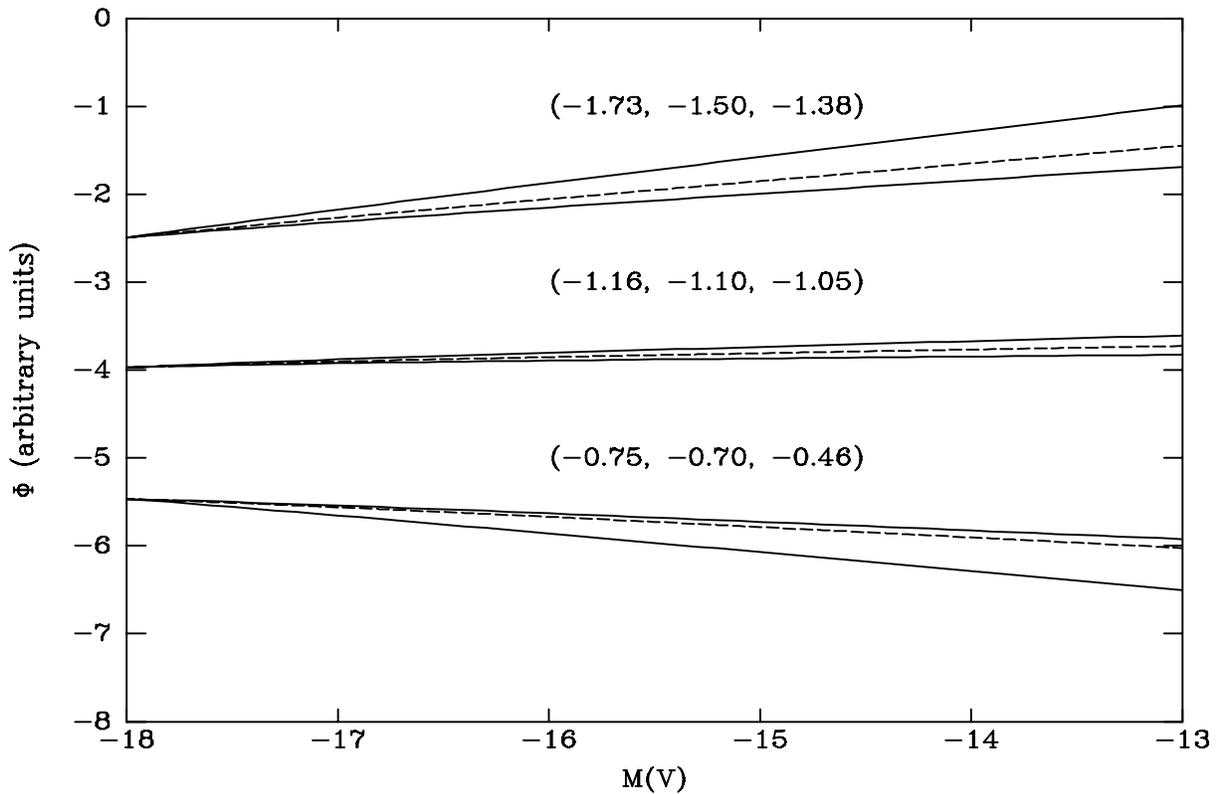}
\caption{
Examples of the results of simulations testing the biasing effect of 
probability-weighted luminosity functions on the measurement of $\alpha$.
The dotted lines represent $\alpha$ of the input LFs, and the solid lines 
show the boundaries wherein 95\% of the output LFs are contained.  The
three numbers above each set of lines are the $\alpha$ of the upper bound, 
input, and lower bound LFs respectively.
}
\label{ALPHAs}
\end{figure}

We generated 150 such simulated LFs for each $\alpha$ tested.  To illustrate
the results of these simulations, we present three sets of them in Figure \ref{ALPHAs}. 
The dashed lines show the input faint-end slopes ($\alpha = -0.7, -1.1,$ and $-1.5$)
for the simulations.  Of the 150 simulations, we exclude the four outliers 
furthest above and four furthest below the input $\alpha$; the remaining 142 simulations
(i.e. 95\%) yield results that fall within the envelope bounded by the solid lines 
above and below each dashed line.  The $\alpha$ represented by each of those solid
and dashed lines is given in the figure.  

As Figure \ref{ALPHAs} shows, the "fuzzing" of the galaxies due to $z_{p}$ uncertainties does cause
systematic biases of the calculated faint-end LF slope.  At $\alpha \simeq -1.1$, the
bias is negligible, and the fuzzing basically just increases the uncertainty 
of the measured slope.  For steeper input $\alpha$, however, the output 
$\alpha$ is quantitatively biased toward steeper values; and
for shallower input $\alpha$, the opposite is true.  The bias increases as input values of
$\alpha$ grow more extreme; an input of $\alpha = -1.9$, for example, produces output
$\alpha$ in a 95\% envelope ranging from $-2.3 \lesssim \alpha \lesssim -1.8$, whereas
an input of $\alpha = -0.4$ produces corresponding output of $+0.1 \lesssim \alpha \lesssim -0.5$.

From all the simulation results, standard bootstrap methods were used to estimate the expected systematic offsets in $\alpha$ for an actually observed galaxy population.  These offsets were then used to correct the computed luminosity functions, to recover the original faint-end slopes
of the galaxy samples in each redshift bin.  This correction affects the overall slope only, and is not intended to remove any inherent, non-power law structure in the observed LF.  Also, this correction strategy was optimized to recover the faint-end slope, rather than the characteristic magnitude $M^{*}$ or the number density normalization $\phi^{*}$ (as was done, for example, in \citealt{che03}).  We thus do not attempt to measure those parameters in this work.  Accurate measurement of those values for the COSMOS survey are presented, for a slightly different galaxy spectral type classification scheme, by \citet{sca07}.

\subsection{Probability-Weighted Luminosity Functions}

\begin{figure}[ht]
\epsscale{1.0} 
\plotone{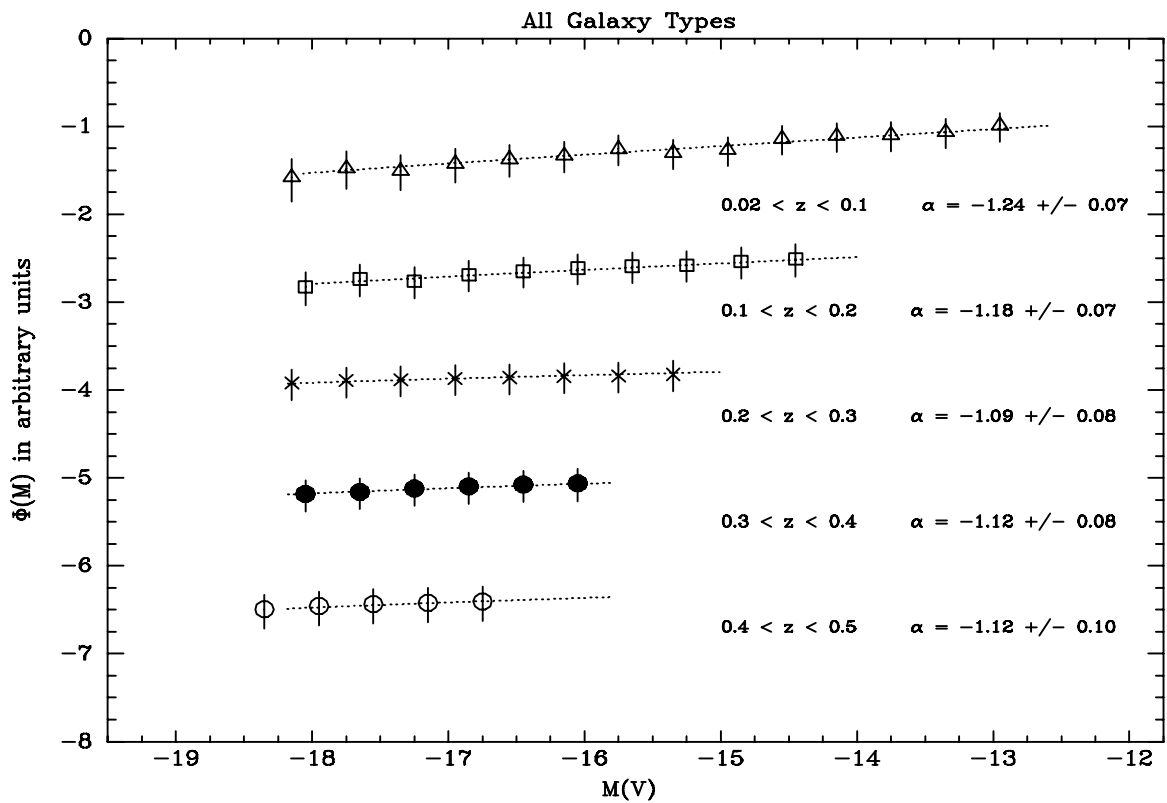}
\caption{
Faint-end portions of the V-band galaxy luminosity functions, $\Phi(M_{V})$, 
in the redshift range $0.02 \leq z \leq 0.5$.
Each LF is offset by a constant for clarity.  The 
best weighted least-squares-fitting faint-end power law 
slope for each LF has been overplotted (dotted lines)
and labeled with its redshift bin and slope.  
} 
\label{LFs_1}
\end{figure}

\begin{figure}[ht]
\epsscale{1.0} 
\includegraphics[scale=0.50]{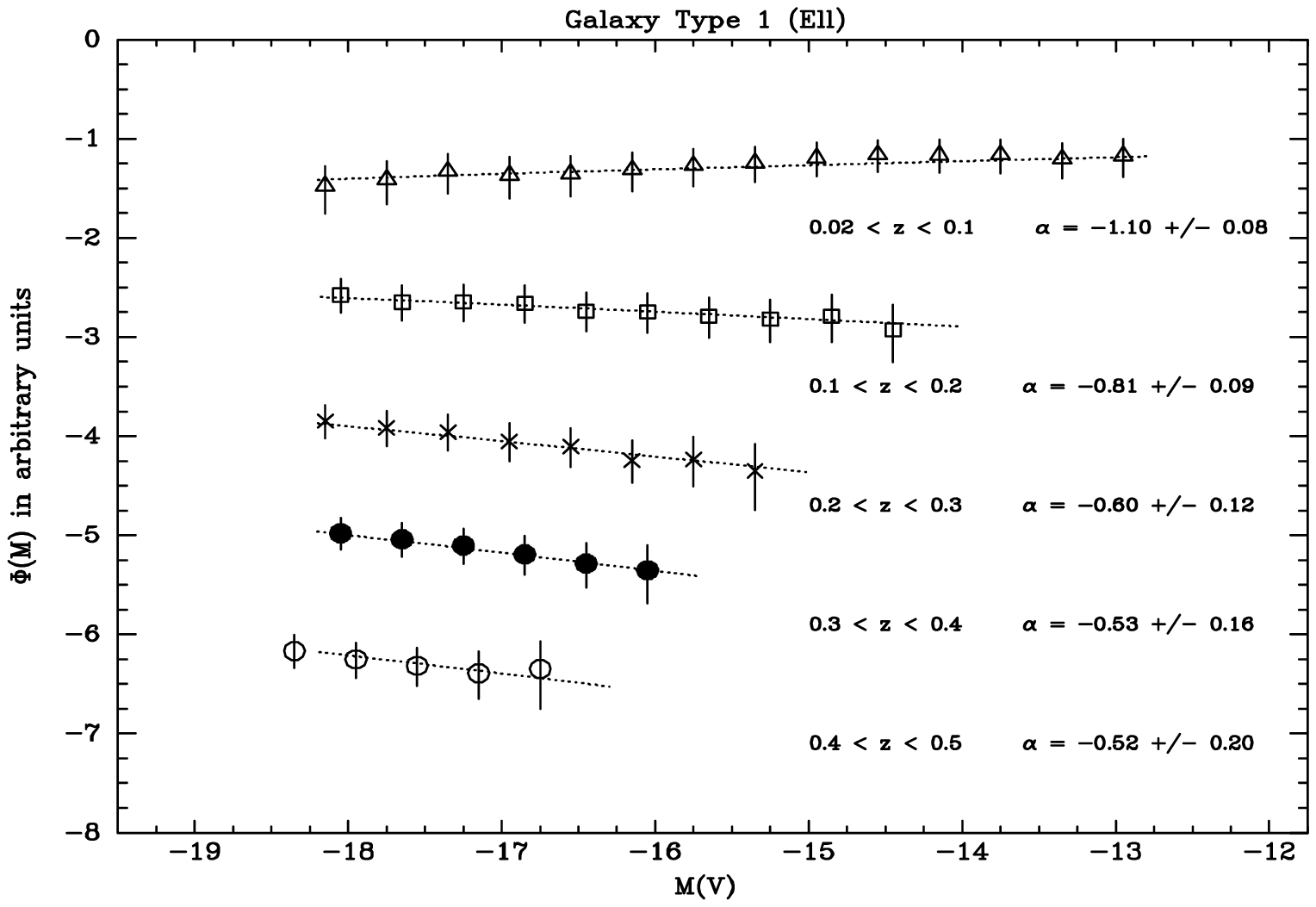}
\includegraphics[scale=0.50]{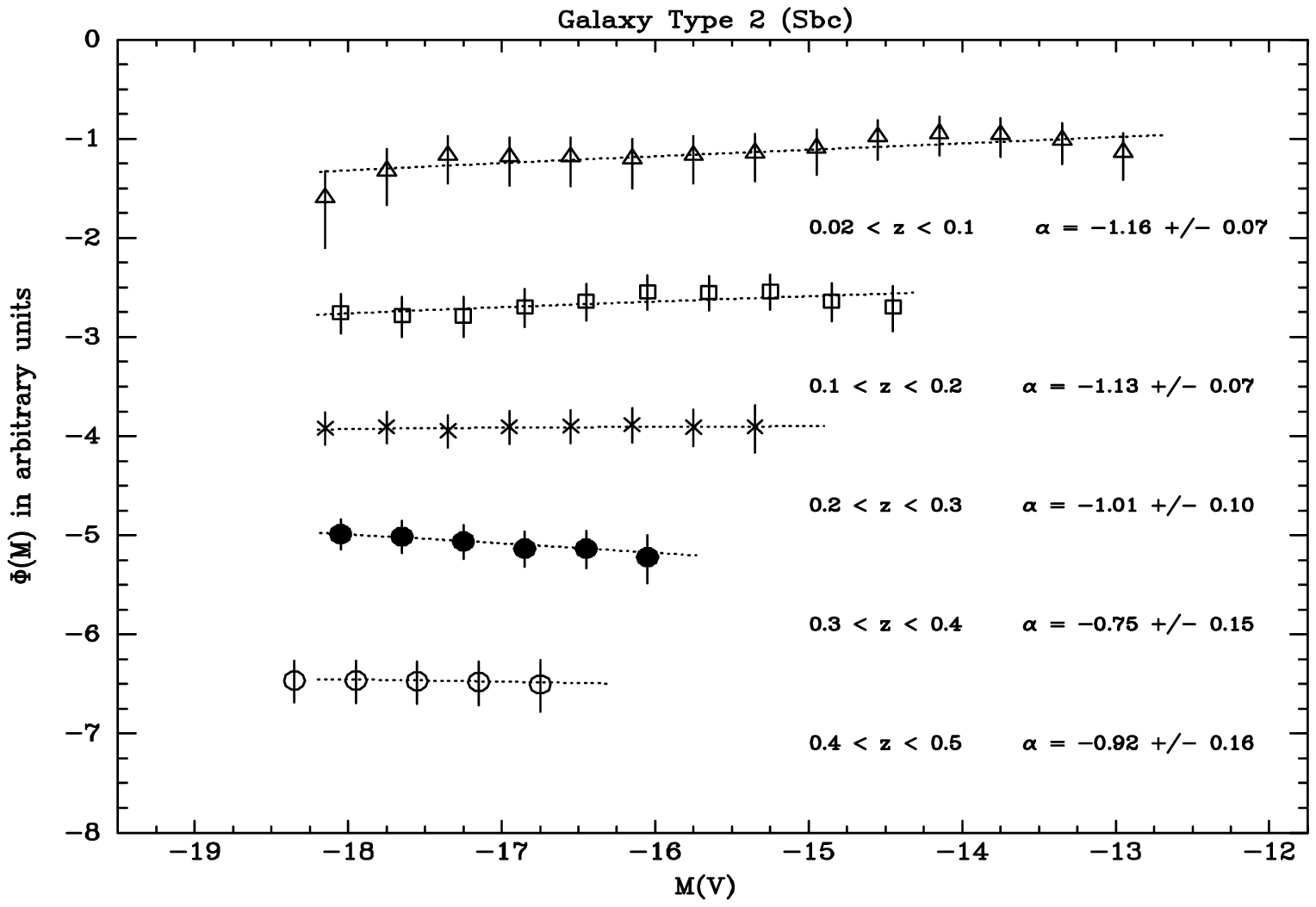}
\includegraphics[scale=0.50]{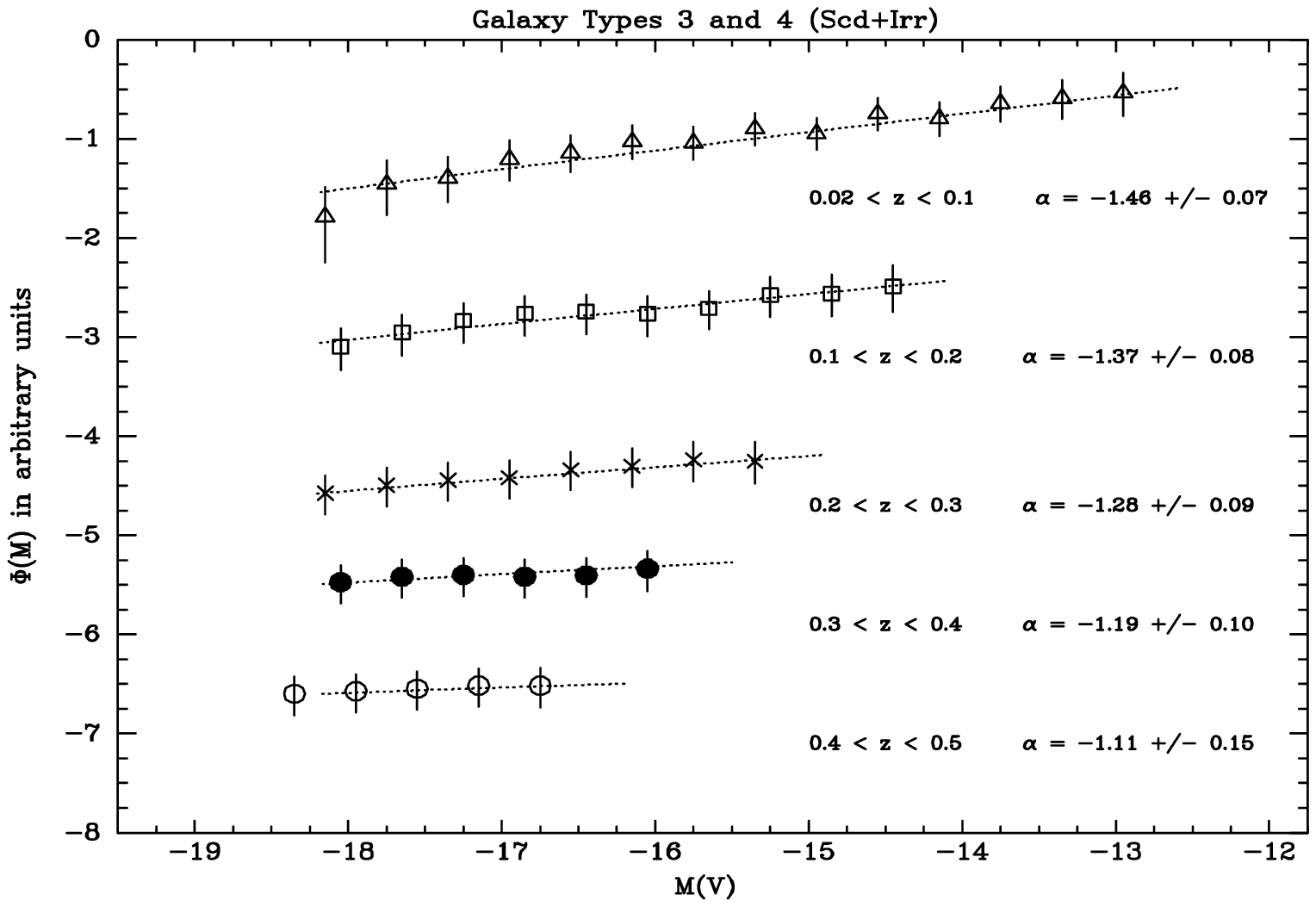}
\includegraphics[scale=0.50]{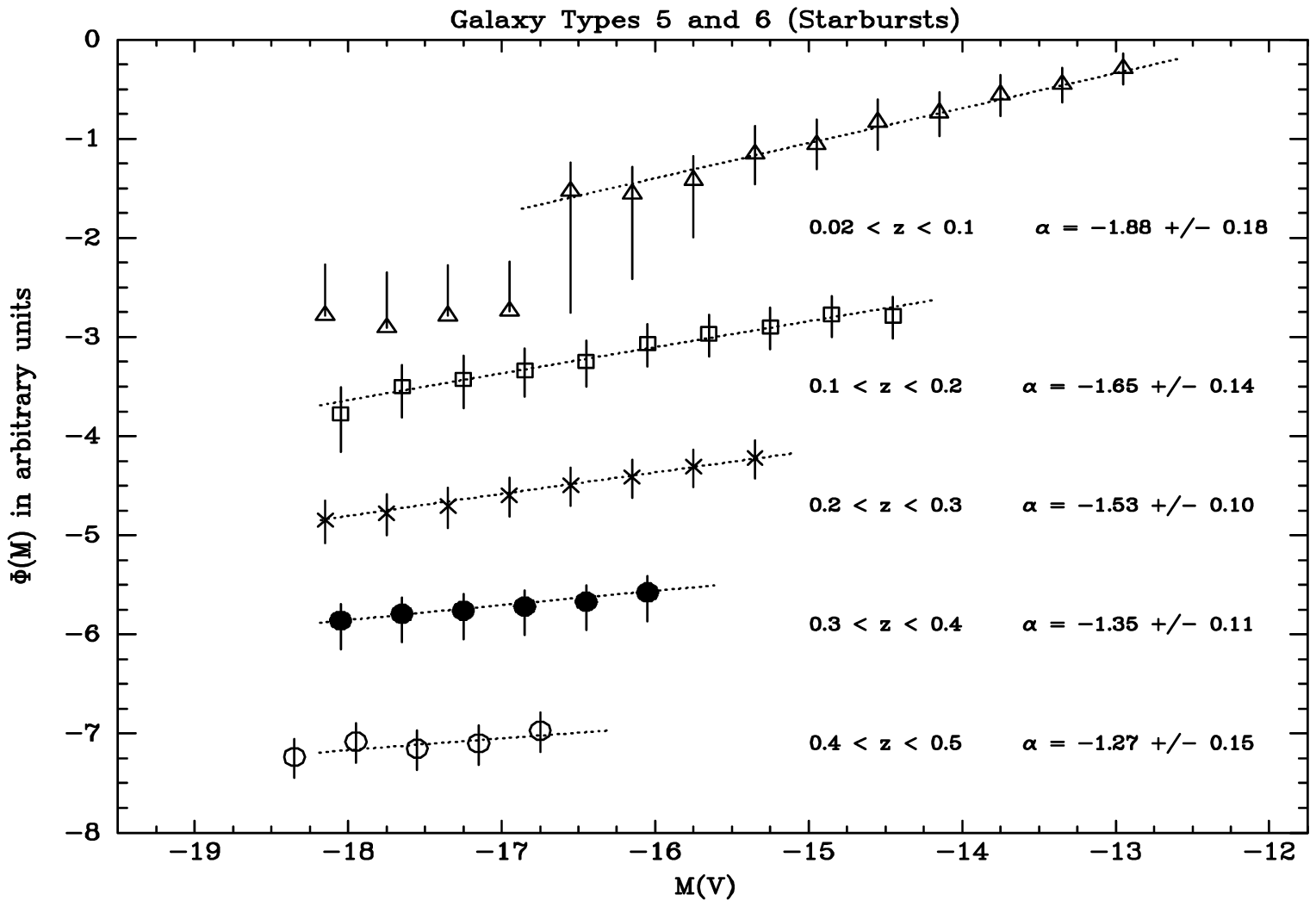}
\caption{
Faint-end portions of V-band luminosity functions, $\Phi(M_{V})$,  for galaxies 
divided by spectral type.  Symbols are the same as in Figure \ref{LFs_1}.  
} 
\label{LFs_abcd}
\end{figure}

The measured luminosity functions for the entire galaxy sample, corrected for this bias in $\alpha$,  are presented in Figure \ref{LFs_1}.  Only the faint ends of the LFs, operationally defined here as galaxies fainter than $M_V \simeq -18$, are presented.  LFs were calculated in five redshift bins: $0.02 < z \leq 0.1$, $0.1 < z \leq 0.2$, $0.2 < z \leq 0.3$, $0.3 < z \leq 0.4$, and $0.4 < z \leq 0.5$.  Additionally, we divided the galaxies into four subsamples according to galaxy spectral type: Type 1 (ellipticals), Type 2 (Sbc), Types 3 and 4 combined (Scd + Irr), and Types 5 and 6 combined (low-extinction starbursts).  For each subsample, LFs were also calculated in the same five redshift bins.  The results are presented in Figure \ref{LFs_abcd}.

There are three primary sources of errors in the LFs: (1) the systematic error in the LF slope and absolute magnitude determinations, described in the text above and characterized using simulations; (2) a Poisson-like error which derives naturally from the modified $1/V_{max}$ method, which is the reciprocal of the square root of the total number of fractional galaxies in each bin; and (3) a non-Gaussian error as a function of absolute magnitude, due to the asymmetric uncertainty in the photometric redshift determination of each galaxy.  The second source of error, because of our large galaxy sample size, is much smaller than the third source of error.  We computed the error from that third source with a standard bootstrap technique, using 150 random samplings (with duplication allowed) of the observed dataset to  determine the 68\% confidence intervals
for $\Phi$ in each magnitude bin.  

Each LF segment was fit to a power-law slope using weighted least squares.  To avoid possible biasing of the faint-end slope by galaxies near $M*$, which as we mentioned above is not well determined by our technique, we only use data fainter than $M_V = -18.3$ in our fits.  In one case, we 
make a more restrictive magnitude cutoff:   the lowest redshift bin in the starburst spectral type.
There, the measured LF drops discontinuously at $M_V = -17$.  We suspect this has occurred 
because $M*$ may be quite faint for this galaxy type and redshift bin, thus distorting 
our measurement of the LF brightward of that point.  So to ensure that 
the $\alpha$ fit to that LF is not correspondingly distorted, we use only data 
fainter than $M_V = -16.8$ for that particular fit.  In both Figures \ref{LFs_1}
and \ref{LFs_abcd}, the measured LF data points are plotted as symbols, and the best-fit $\alpha$
values are plotted as dotted lines.  The faint-end slope fits are summarized in Table 2.

\section{Discussion}

Because of the combined large area and depth of the COSMOS survey, the 
luminosity functions presented in this work provide a glimpse of $\alpha$
across a substantial range of redshift in a single, consistent dataset.  
With this view, our results show that for all spectral types combined,
$\alpha = 1.24 \pm 0.07$ for the local ($0.02 < z \leq 0.1$) universe.
As the redshift increases, $\alpha$ flattens out somewhat, and is
$-1.12 \pm 0.10$ in our highest redshift bin ($0.4 < z \leq 0.5$).  

Our local LF is consistent with results from the two largest local galaxy 
surveys to date, which have comparable ($\sim 10^5$ galaxies)
sample sizes to our study here.  The 2dFGRS survey found that, for the
$b_J$ band galaxy luminosity function, $\alpha = -1.21 \pm 0.03$ (\citealt{nor02}) 
and $-1.18 \pm 0.02$ for the redshift range $0.02 < z < 0.25$ (\citealt{cro05}).
The SDSS has $\alpha = -1.05 \pm 0.01$ for
a slightly redshifted $r$-band galaxy LF for galaxies brighter than $M_r \sim -17$ 
(\citealt{bla03}), and $\alpha \simeq -1.3$ at fainter magnitudes (\citealt{bla05}).
\citet{bla05} has further shown that, with the appropriate conversion 
of the 2dFGRS $b_J$ data, it and the SDSS $g$-band LFs have consistent 
low-luminosity slopes. 
At higher redshift, our results are consistent with \citet{sca07}, who independently
derived $\alpha = -1.26 \pm 0.15$ in the range $0.2 < z < 0.4$ 
for a portion of the COSMOS survey area.
Our results are also consistent with results from other surveys given in the 
literature - for example, with the V-band LF derived from the VVDS (\citealt{ilb05}), 
where $\alpha = 1.21 \pm 0.04$ in the range $0.2 < z < 0.4$.

The formal errors in our $\alpha$ measurements are higher than those
of most of these other studies; this is probably mainly because of the 
systematic slope uncertainties that we attempt to account for with our simulations.
Our overall agreement, 
however, appears to confirm that we have properly accounted for the errors
that result from representing galaxies as probability-smoothed 
luminosity distributions.

\subsection{LFs as a Function of Galaxy Spectral Type}

The luminosity functions in the four galaxy spectral type bins we used - SED
templates of early-type, Sbc, Scd+Irr, and low-extinction starbursts - follow the 
well-known pattern of steeper $\alpha$ for bluer galaxies.  In our low-redshift
bin, $\alpha$ increases from $-1.10 \pm 0.08$ in early-types to $-1.88 \pm 0.18$
in starbursts; the trend continues with increasing redshift, showing a similar
steepening of $\alpha$, from $-0.52 \pm 0.20$ to $-1.27 \pm 0.15$.  

As with the full galaxy sample, these type-specific results are
consistent with the findings of previous work in the literature.  Of course,
due to differing galaxy selection criteria and redshift binning, exact comparisons
are not always possible.  Generally speaking, however, for local
galaxies, our red/early-type and intermediate spiral galaxy LFs are consistent
with SDSS and 2dFGRS results; and previous samples of $z \lesssim 0.1$ 
galaxy populations have also shown very steep
$\alpha$ for the bluest and most irregular galaxies: $\alpha = -1.87$ for the
CfA Redshift Survey (\citealt{mar94}), $-1.84$ for the Las Campanas Redshift
Survey (\citealt{bro98}), $-1.81$ for the SSRS2 (\citealt{mar98}), and $-1.9$
from the Deep Multicolor Survey (\citealt{liu98b}) and SDSS (\citealt{nak03}).
At higher redshifts, comparisons with COMBO-17 (\citealt{wol03}), VVDS
(\citealt{zuc06}), and with the COSMOS survey itself (\citealt{sca07}) show
broad consistency across the various galaxy type and redshift intervals.

\subsection{$\alpha$ vs. $z$: Evolution Or Selection?}

In the context of the broad consistency of our results with those in the literature,
perhaps the most striking result in this work is the clear trend, with every 
galaxy spectral type, of a flattening of the faint-end slope with increasing redshift.
From our lowest redshift bin to our highest - i.e., from $z \sim 0$ to $z \sim 0.5$ - 
$\Delta \alpha = 0.58, 0.24, 0.35,$ and $0.61$ respectively for early-types,
Sbc, Scd+Irr, and low-extinction starbursts.

On the surface, this trend may not appear to be consistent with previous work.
Much of the work to derive the evolution of galaxy LF parameters $M*$ and $\Phi*$
as a function of redshift (e.g., \citealt{lin99,wol03,bal05,wil06}),
in fact, depends upon the assumption that $\alpha$ does not evolve with redshift,
or at most weakly evolves to $z \lesssim 1$.  We have focused here on the 
measurement of $\alpha$ rather than those other parameters; that, and the
fact that all of our measurements have come from a single dataset, plus the fact
that each individual determination of $\alpha$ is consistent with previous work,
supports the likelihood that this observed flattening trend is real.

The question is, do these changing slopes represent true evolution 
in $\alpha$, or do they reflect our ability to detect different galaxy
populations as a function of redshift?  The latter possibility can be discussed
in the context of, among others, \citet{del03} and \citet{bla05}, who suggest
that the faint end of the field galaxy LF is comprised of a composite population
of dwarf and non-dwarf galaxies, each with its own functional form.  This
would mean that a single power law is not quite sufficient to describe the
LF faint end accurately.  \citet{bla05} further suggests, through detailed 
examination of the faint galaxy population in the SDSS, that a large fraction of
these dwarf galaxies may have very low surface brightness, and are thus not 
included in most faint end LF measurements.

Our dependence on photometric redshifts places an important caveat
on the interpretation of our data: by using "fuzzy" galaxies, any second-order
deviations from a power law at the faint end of the LF may well have been
smoothed out, and are thus not recoverable from our LF measurements.
So if a deviation from a single faint-end power law does occur at very low 
luminosities, we cannot address that issue with this work. 

Due to the substantial depth of the COSMOS survey imaging, 
it is likely that we have successfully measured a larger fraction of
low surface brightness dwarf galaxies than wider-area, shallower surveys such as
SDSS or 2dFGRS.  The steepness of our low-redshift LFs may reflect this.
Even the COSMOS survey depth, however, does not allow us to measure $\alpha$
fainter than $M_V \sim -17$ at $z = 0.5$, so we cannot say if the flattening
of $\alpha$ in our higher redshift bins is due to the non-detection of these
dwarfs.  There may be some circumstantial evidence, however, to support
that picture.  For example, \citet{dah05} measured for the GOODS survey
$\alpha = -1.37$ for the rest-frame B-band from
$0.1 < z < 0.5$, which is somewhat steeper than most LF measurements
in this range.  However, the GOODS survey is very deep; so in this broad
redshift bin, a large detected fraction of faint, low surface brightness dwarfs
near $z \gtrsim 0.1$ could have driven $\alpha$ to a steeper value for the full range.

As the spectroscopic portion of the COSMOS survey (\citealt{lil07}) is 
completed, we will be able to address this question in more detail, as we 
deconvolve the faint end of the galaxy LF as a multivariate function of color, 
morphology, luminosity, and redshift.

\section{Conclusions}

Using the COSMOS multiband photometry and photometric redshift catalog,
we have constructed faint-end rest-frame V-band luminosity functions for the galaxy
population at $0.02 < z \leq 0.5$ in the COSMOS survey volume.
Since we are using photometric redshifts, we have computed these LFs
by treating galaxies as weighted probability-smoothed luminosity distributions,
and using a modified $1/V_{max}$ method.  A total of 49,161 galaxies have
photometric redshifts that fall in this redshift range; within and outside this
range, a total of 80,820 galaxies contribute to the derived LFs.  Extensive
Monte Carlo simulations were used to characterize and account for
the systematic and random errors of this technique. 

For all galaxy spectral types, the LF slope ranges from $-1.24$ to $-1.12$ 
from the lowest redshift bin to the highest. 
In the lowest redshift bin ($0.02 < z < 0.1$), where the magnitude limit is
$M_{V} \lesssim -13$, the slope ranges from $\alpha = -1.10 \pm 0.08$ for
galaxies with early-type spectral energy distributions (SEDs), to
$\alpha = -1.88 \pm 0.18$ for galaxies with low-extinction starburst SEDs.
In each galaxy SED category (early-type, Sbc, Scd+Irr, and starburst), the
faint-end slopes grow shallower with increasing redshift; in the highest
redshift bin ($0.4 < z < 0.5$),
$\alpha = -0.52 \pm 0.20$ and $-1.27 \pm 0.15$ for early-types 
and starbursts respectively.  

All of our derived type-specific LFs, across our redshift ranges,
are broadly consistent with the findings of previous authors.  
Our results thus show a flattening trend for $\alpha$ with increasing
redshift for each spectral type.  It is unclear, however, if this is evidence
of $\alpha$ evolution in the galaxy LF, or of preferential selection of
dwarf galaxies in the local universe.  The steepness of $\alpha$ at
lower redshift could be qualitatively explained, for example, by large numbers of faint
dwarfs, perhaps of low surface brightness, which are not detected
at higher redshifts.  We will address this question in a future paper, as 
the full set of COSMOS data - in particular, spectroscopic redshifts - are
obtained.
 
 \acknowledgments
 
The HST COSMOS Treasury program was supported through NASA grant
HST-GO-09822. We wish to thank Tony Roman, Denise Taylor, and David 
 Soderblom for their assistance in planning and scheduling of the extensive COSMOS 
 observations.  We gratefully acknowledge the contributions of 
 the entire COSMOS collaboration, which contains more than 70 scientists. 
 More information on the COSMOS survey is available \\ at
  {\bf \url{http://www.astro.caltech.edu/~cosmos}}. It is a pleasure to 
 acknowledge the excellent services provided by the NASA IPAC/IRSA 
 staff (Anastasia Laity, Anastasia Alexov, Bruce Berriman and John Good) 
 in providing online archive and server capabilities for the COSMOS datasets.
 The COSMOS Science meeting in May 2005 was supported in part by 
 the NSF through grant OISE-0456439.  We thank Paris Bogdanos and James Cohen for 
 image processing and data formatting assistance.  C. Liu acknowledges the hospitality
 and support of the Hayden Planetarium and Department of Astrophysics at
 the American Museum of Natural History.  C. Liu, T. Paglione, and S. Tribiano 
 gratefully acknowledge support from a City University of New York CCIR Grant.
 
 {\it Facilities:} \facility{HST (ACS)}, \facility{Subaru}, \facility{KPNO},
 \facility{CTIO}, \facility{CFHT}.
 

\clearpage
 
\begin{deluxetable}{cccccc}
\tabletypesize{\scriptsize}
\tablecaption{Galaxy samples and $z_p$ confidence levels \tablenotemark{a} \label{tbl-1}}
\tablewidth{0pt}
\tablehead{
   \colhead{\# Galaxies  }	&  
   \colhead{Redshift constraint}	& 
   \colhead{$\delta_{68}$ constraint\tablenotemark{b}}	& 
   \colhead{median $\delta_{68} / (1+z) $}	&
   \colhead{mean $\delta_{68} / (1+z)$}	& 
   \colhead{rms $\delta_{68} / (1+z) $}
   }
 \startdata

41237	& $-$					& $\delta_{68} \leq 0.1$ & 0.042	& 0.041	& 0.053	 \\
49161	&  $0.02 \leq z \leq 0.50$	&  $-$				& 0.043	& 0.050	& 0.073	 \\ 
80280	& $-$					&  $-$				& 0.052	& 0.081	& 0.239	 \\
 
\enddata 
\tablenotetext{a}{Galaxy spectral types as defined by \citet{mob07}}
\tablenotetext{b}{$\delta_{68} = $ the width of the 68\% confidence interval for $z_p$.}
\end{deluxetable}

 \clearpage
 
 \begin{deluxetable}{lcr}
 \tabletypesize{\scriptsize}
 \tablecaption{Luminosity Function Slope Fits \label{tbl-2}}
 \tablewidth{0pt}
 \tablehead{
 \colhead{Galaxy Spectral Type \tablenotemark{a} } &  \colhead{Redshift Range} & \colhead{Slope $\alpha$}  } 
\startdata
T1-T6 (All)			&  $0.02 \leq z < 0.1$	&  $-1.24 \pm 0.07$ \\ 
 					&  $0.1 \leq z < 0.2$		&  $-1.18 \pm 0.07$ \\ 
 					&  $0.2 \leq z < 0.3$		&  $-1.09 \pm 0.08$ \\ 
 					&  $0.3 \leq z < 0.4$		&  $-1.12 \pm 0.08$ \\ 
 					&  $0.4 \leq z < 0.5$		&  $-1.12 \pm 0.10$ \\ 
					
T1 (Ell)				&  $0.02 \leq z < 0.1$	&  $-1.10 \pm 0.08$ \\ 
 					&  $0.1 \leq z < 0.2$		&  $-0.81 \pm 0.09$ \\ 
 					&  $0.2 \leq z < 0.3$		&  $-0.60 \pm 0.12$ \\ 
 					&  $0.3 \leq z < 0.4$		&  $-0.53 \pm 0.16$ \\ 
 					&  $0.4 \leq z < 0.5$		&  $-0.52 \pm 0.20$ \\ 
					
T2 (Sbc)				&  $0.02 \leq z < 0.1$   	&  $-1.16 \pm 0.07$ \\ 
 					&  $0.1 \leq z < 0.2$		&  $-1.13 \pm 0.07$ \\ 
 					&  $0.2 \leq z < 0.3$		&  $-1.01 \pm 0.10$ \\ 
 					&  $0.3 \leq z < 0.4$		&  $-0.75 \pm 0.15$ \\ 
 					&  $0.4 \leq z < 0.5$		&  $-0.92 \pm 0.16$ \\ 

T3+T4 (Scd+Irr)		&  $0.02 \leq z < 0.1$   	&  $-1.46 \pm 0.07$ \\
					&  $0.1 \leq z < 0.2$		&  $-1.37 \pm 0.08$ \\ 
 					&  $0.2 \leq z < 0.3$		&  $-1.28 \pm 0.09$ \\ 
 					&  $0.3 \leq z < 0.4$		&  $-1.19 \pm 0.10$ \\ 
 					&  $0.4 \leq z < 0.5$		&  $-1.11 \pm 0.15$ \\
					 
T5+T6 (Starbursts)		&  $0.02 \leq z < 0.1$   	&  $-1.88 \pm 0.18$ \\
 	 				&  $0.1 \leq z < 0.2$		&  $-1.65 \pm 0.14$ \\ 
 					&  $0.2 \leq z < 0.3$		&  $-1.53 \pm 0.10$ \\ 
 					&  $0.3 \leq z < 0.4$		&  $-1.35 \pm 0.11$ \\ 
 					&  $0.4 \leq z < 0.5$		&  $-1.27 \pm 0.15$ \\ 
\enddata
\tablenotetext{a}{Galaxy spectral types as defined by \citet{mob07}}
\end{deluxetable}

\clearpage

\clearpage

\clearpage

\clearpage

\clearpage

 
 \end{document}